\documentclass[iop]{emulateapj}
\usepackage{epsfig, amsmath, natbib,graphicx}

\begin{document}

\title{The effects of patchy reionization on satellite galaxies of the Milky Way}
\submitted{ApJ in press}
\author{Ragnhild Lunnan$^1$, Mark Vogelsberger$^1$, Anna Frebel$^1$, Lars Hernquist$^1$, Adam Lidz$^2$ and Michael Boylan-Kolchin$^{3}$}
\affil{$^1$ Harvard-Smithsonian Center for Astrophysics, 60 Garden Street, Cambridge, MA 02138}
\affil{$^2$ Department of Physics \& Astronomy, University of Pennsylvania, 209 South 33rd Street, Philadelphia, PA 19104}
\affil{$^3$ Center for Galaxy Evolution, 4129 Reines Hall, University of California, Irvine, CA 92697}
\email{rlunnan@cfa.harvard.edu}

\begin{abstract}

We combine the high-resolution Aquarius simulations with three-dimensional models of reionization based on the initial density field of the Aquarius parent simulation, Millennium-II, to study the impact of patchy reionization on the faint satellite population of Milky Way halos. Because the Aquarius suite consists of zoom-in simulations of halos in the Millennium-II volume, we follow the formation of substructure and the growth of reionization bubbles due to the larger environment simultaneously, and thereby determine the reionization redshifts of satellite candidates. We do this for four different reionization models, and also compare results to instantaneous reionization. Using a simple procedure for selecting satellites and assigning luminosities in the simulations, we compare the resulting satellite populations. We find that the overall number of satellites depends sensitively on the reionization model, with a factor of 3-4 variation between the four models for a given host halo, although the difference is entirely in the population of faint satellites ($M_V > -10$). In addition, we find that for a given reionization model the total number of satellites differs by 10-20\% between the patchy and homogeneous scenarios,
provided that the redshift is chosen appropriately for the instantaneous case. However, the halo-halo scatter from the six Aquarius halos is large, up to a factor of 2-3, and so is comparable to the difference between reionization scenarios. In order to use the population of faint dwarf galaxies around the Milky Way as a probe of the local reionization history, then, it is necessary to first better understand the general distribution of substructure around Milky Way-mass halos. 

\end{abstract}



\section{Introduction}\label{sec:intro}

A major challenge to the $\Lambda$CDM cosmological model is its
over-prediction of substructure on small scales compared to the
actual observed number of dwarf galaxies in the Local Group (the
so-called ``missing satellite problem''; \citealt{Moore1999, Klypin1999}).
The most well-studied explanation for this discrepancy is that the lowest
mass halos were inefficient at forming stars, so that most of these
substructures remain dark.  Indeed, several processes can suppress star
formation and reduce the number of visible halos, including tidal
stripping of satellites \citep[e.g.,][]{Kravtsov2004b}, supernova
feedback, and the increase of gas temperature due to cosmic
reionization \citep{Efstathiou1992, Gnedin2000, Benson2002b,
Somerville2002, Madau2008b}.  Dynamical interactions between
dwarfs \citep{Donghia2009} or between dwarfs and the luminous
disks of large halos \citep{Donghia2010} may also play a role.

The discovery of a population of ``ultra-faint'' dwarf galaxies (UFDs;
$L < 10^5 L_{\odot}$) in the Sloan Digital Sky Survey (SDSS)
\citep[e.g.,][]{Willman2005, Belokurov2006, Zucker2006b} has roughly
doubled the number of known satellites, alleviating the problem
somewhat.  Based on the sky coverage and sensitivity of SDSS, estimates
of a completeness-corrected luminosity function down to $M_V \sim -2$
\citep{Koposov2008, Tollerud2008} are available.  The possible number
of faint satellites could be up to the order of hundreds, but due to
the incomplete sky coverage of SDSS, the exact number remains uncertain.

The discovery of these low-luminosity dwarfs has also raised
many new questions, however. Follow-up observations have shown that they are
the faintest, most dark matter-dominated and metal-poor systems known
\citep{Simon2007, Kirby2008b}, with metallicity patterns similar to
the lowest metallicity stars in the halo \citep[e.g.,][]{Frebel2010a}.
In order to understand the cosmological context of UFDs, including
their relationship to the first galaxies, and to the building blocks
of the stellar halo, it is important to be able to model conditions in
the universe at the time of their formation (e.g., \citealt{Frebel2010c}).

Semi-analytic models of galaxy formation have been able to explain the
observed number of satellites by invoking various feedback mechanisms,
including an external reionization field \citep[e.g.,][]{Benson2002a,
Maccio2010, Li2010}.  According to this scenario, 
as the intergalactic
medium is heated by reionization, 
gas can no longer be accreted by and cool within
halos below a certain virial
temperature,
inhibiting further
star formation in the lowest mass halos.  
However, these models generally
rely on a simplified, uniform reionization field, which is likely
quite different from reality.  Simulations of reionization indicate
that it was likely extended in time and spatially inhomogeneous, with a wide
range in reionization redshifts for different parts of the universe
\citep{Barkana2004, Lidz2007, McQuinn2007, Ahn2009, Alvarez2009}.

Several recent studies have examined various aspects of inhomogeneous
(patchy) reionization and its consequences for galaxy formation.
\citet{Munoz2009} combined a reionization model with merger trees
drawn from the Via Lactea II simulation, and used the observed
properties of satellites to constrain model parameters.
\citet{Alvarez2009} combined N-body and three-dimensional reionization
calculations to investigate the relationship between reionization
history and local environment.  \citet{Busha2010} used their resulting
distribution of reionization redshifts together with the Via Lactea II
subhalos, and found that the resulting number of Milky Way satellites
can vary by an order of magnitude.  Understanding the connection
between environment, reionization epoch, and faint satellites, then,
can in principle help us to understand how the Milky Way formed.

Our work is similar in spirit to the above efforts, in that we aim to
examine how the history of reionization impacts the abundance of faint
satellites around Milky Way-like halos.  However, our methods differ
from the earlier studies in that we combine reionization calculations
with high resolution zoom-in simulations of Milky Way-sized halos {\it
in the same box}, enabling us to track the growth of reionization
bubbles and dark matter substructure simultaneously.  This means that,
subject to the approximations inherent to our treatment,
the reionization history is completely 
determined by the large-scale
environment, and infalling subhalos can have different
reionization redshifts than the Milky Way-like final host.  Moreover,
since we have six high resolution halos available, we can roughly estimate the 
consequences of cosmic variance, rather than assuming that a single dark
matter halo is representative of the Milky Way.  Comparing the results
from the patchy scenarios with instantaneous and homogeneous reionization models, we
aim to quantify how much of a difference ignoring the patchiness of
reionization makes, and to what extent the faint satellites trace the
local reionization epoch.

This paper is organized as follows: Section \ref{sec:sims} describes
the simulations that are the basis of our analysis: the Aquarius suite
of Milky Way halos, and reionization calculations based on its
``parent'' simulation, Millennium II.  Our approach is described in
Section \ref{sec:model}, detailing how we pick out satellite
candidates from the merger trees, and assign luminosities to them.
Section \ref{sec:res} describes our results, showing luminosity functions in different scenarios, and exploring the differences between the patchy and homogeneous models. Our conclusions are summarized
in Section \ref{sec:conc}.


\section{Simulation Data}\label{sec:sims}
\subsection{The Aquarius simulations}\label{sec:aqua}

Our modeling is based on the
Aquarius\footnote{http://www.mpa-garching.mpg.de/aquarius/}
simulations \citep{Springel2008}, a suite of six highly resolved Milky
Way sized dark matter halos.  These are re-simulations of six halos from a lower resolution version ($900^3$ particles) of the Millennium-II simulation, which is a periodic box of size 100$h^{-1}$ Mpc containing $2160^3$ particles \citep{Boylan-Kolchin2009}. The six
halos were selected on the basis of final mass, and by not having a
massive close neighbor at $z=0$ (no late-time major merger); otherwise
they were selected randomly.  We note that one of these halos (the F
halo) is somewhat of an outlier in terms of its merger history, in
that it experienced its last major merger at $z=0.7$.

The same cosmological parameters were adopted as for the original
Millennium simulation \citep{Springel2005}: $\Omega_m = 0.25$,
$\Omega_{\Lambda} = 0.75$, $\sigma_8 = 0.9$, $n_s = 1$ and Hubble
constant $H_0 = 100 h$ km s$^{-1}$ Mpc$^{-1}$ = 73 km s$^{-1}$
Mpc$^{-1}$.  While these are consistent within the uncertainties with
the parameters estimated from the three-year Wilkinson Microwave Anisotropy Probe (WMAP) data
\citep{Spergel2007}, the value of $\sigma_8 = 0.9$ is slightly higher
than that inferred from the most recent measurements
\citep{Komatsu2011}.  In principle, the adopted $\sigma_8$ could
lead to an overestimate of the number of halos collapsing around the
reionization epoch, but studies like \citet{Boylan-Kolchin2010b} suggest this would 
have a relatively minor effect.

All six halos (labeled `A' to `F') were simulated with at least two
different resolutions; we use the higher of the two (``level 2'';
softening length of 65 pc) in our analysis.  Table~\ref{tab:aq_halos}
lists the particle mass, final halo mass, and virial radius of each of
the halos.  Mass and radius are listed as $M_{200}$ and $r_{200}$,
defined as the mass enclosed in a sphere with mean density 200 times
the critical density, and the corresponding virial radius.

Snapshots are stored at 128 output times, from redshift 127 to 0,
equally spaced in $\log a$ where $a = 1/(1+z)$.  For each snapshot,
subhalos are identified using the {\small SUBFIND} algorithm
\citep{Springel2001}, and linked between snapshots in a merger tree.
Since {\small SUBFIND} requires a minimum of 20 gravitationally bound
particles to define a subhalo, the smallest halo masses we resolve are
of order a few times $10^5$ solar masses.  The merger trees are our
starting point for identifying subhalos in the simulation that may
host present-day luminous satellite galaxies.

\begin{table}
 \begin{center}
   \caption{Basic parameters of the six Aquarius halos.}
   \label{tab:aq_halos}
   \begin{tabular}{c c c c}
     \hline \hline
     Name   & $m_p$ $(M_{\odot})$  & $M_{200}$ $(M_{\odot})$ & $r_{200}$ (kpc) \\ \hline 
     Aq-A-2 & $1.370 \times 10^4$  & $1.842 \times 10^{12}$  &  245.88         \\
     Aq-B-2 & $6.447 \times 10^3$  & $8.194 \times 10^{11}$  &  187.70         \\
     Aq-C-2 & $1.399 \times 10^4$  & $1.774 \times 10^{12}$  &  242.82         \\
     Aq-D-2 & $1.397 \times 10^4$  & $1.774 \times 10^{12}$  &  242.85         \\
     Aq-E-2 & $9.593 \times 10^3$  & $1.185 \times 10^{12}$  &  212.28         \\
     Aq-F-2 & $6.776 \times 10^3$  & $1.135 \times 10^{12}$  &  209.21         \\ \hline
    
  \end{tabular}
 \end{center}
\end{table}

\subsection{Reionization calculations}\label{sec:reion_cal}

The Aquarius halos are drawn from a larger parent simulation, so we
can use the parent box to calculate how regions around these halos
reionize 
depending on their environment.  We use the semi-analytic approximation of \citet{Zahn2007},
based on the 
excursion set treatment of \citet{Furlanetto2004a,
Furlanetto2004b}. This semi-analytic model assumes that galaxies reside in halos above some minimum mass, $M_{\rm min}$,
and that the number of ionizing photons produced by a galaxy is directly
proportional to its host halo mass. A given region is considered ionized when the collapse fraction
-- i.e., the fraction of matter that lies in halos above the minimum mass -- in the region
exceeds some threshold value. This threshold depends on the efficiency of the ionizing sources,
and is smaller for models with more efficient sources. For the purpose of modeling reionization, we
fix the minimum mass at $M_{\rm min} = 10^8 M_\odot$, and consider several models for the source efficiencies,
spanning a range of possibilities for the timing and duration of reionization.

\begin{figure}
 \begin{center}
  \includegraphics[width=8.5cm]{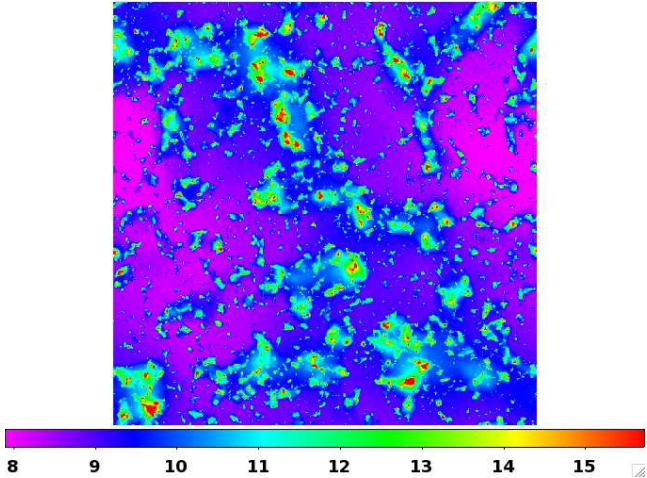}
 \end{center}
 \caption{Slice through a reionization map from Model 1, illustrating the patchy nature of the process. The color indicates the redshift of reionization. The slice is 100 $h^{-1}$ Mpc on a side, and 195 $h^{-1}$ kpc (the cell size in the reionization calculations) thick.}
 \label{fig:reion_map}
\end{figure}

 We take the initial conditions of the Millennium-II
simulation as input to generate reionization maps for the cube of
100 $h^{-1}$ Mpc per side.  Reionization redshifts are tagged spatially, at
a resolution of 512 cells per side, or about 195 $h^{-1}$ kpc.  As an
example, a slice through one of the reionization cubes is shown in
Fig.~\ref{fig:reion_map}, illustrating the patchiness of the process.\footnote{The reionization maps are available from the first author upon request.}

\begin{figure}
 \begin{center}
  \includegraphics[width=8.5cm]{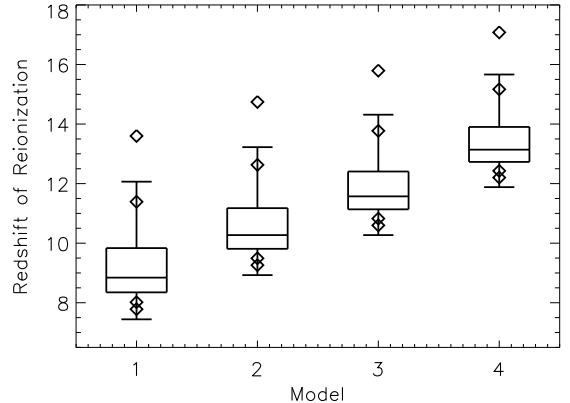} 
 \end{center}
 \caption{Boxplot summarizing the distributions of reionization redshifts over the cells in the 100 $h^{-1}$ Mpc box, for the four different models we considered. The boxes enclose the interquartile range, and the horizontal line is the median in each case. The whiskers extend to the highest and lowest value still within $1.5 \times$ the interquartile range, and the 2nd, 9th, 91st and 98th percentile are marked by diamonds.}
 \label{fig:reion_bp}
\end{figure}

We calculate four different models, including a very efficient
reionization that completes around redshift 12, and three gradually
more extended models.  The distributions of reionization redshifts in
each model, over the cells in the box, are summarized in
Table~\ref{tab:reion_mod} and Fig.~\ref{fig:reion_bp}. 
In the Table, $<z_{reion}>$
and $\sigma_z$ refer to the mean and standard deviation of the
reionization redshifts across the cells in the box, and $z_{end}$ is
the redshift where the process is complete.  
The quantity $\tau_e$ is the Thomson
optical depth; the 7-year WMAP value for this is $\tau_e = 0.088 \pm
0.015$ \citep{Komatsu2011}.

With these maps, we can then find the reionization redshift of any
given subhalo as follows.  At each snapshot, we locate the position of
the subhalo in the parent box, and look up the redshift at which this
region of the parent box reionizes.  The first snapshot where this
redshift is earlier than the redshift of the snapshot, 
flags the subhalo as having
entered a reionized region, and we define this as the reionization
redshift of the subhalo. Note that this allows not only
for 
different reionization redshifts for the six main Aquarius halos, but also that their subhalos ending up as satellites are not required to reionize at the same time as the host.

\begin{table}
 \begin{center}
  \caption{Properties of the reionization models}
  \label{tab:reion_mod}
  \begin{tabular}{c c c c c c}
   \hline \hline
   Model No. &  $<z_{reion}>$   & $\sigma_z$ & $z_{end}$       & Median        & $\tau_e$    \\ \hline
   1         & $\mbox{ }9.31$   &  1.44      &  $\mbox{ }7.44$ & $\mbox{ }8.84$ & 0.075       \\
   2         & 10.71            &  1.35      &  $\mbox{ }8.92$ &  10.27         & 0.091       \\
   3         & 11.97            &  1.28      &  10.27          &  11.57         & 0.108       \\
   4         & 13.50            &  1.20      &  11.88          &  13.14         & 0.129       \\ \hline
  \end{tabular}
 \end{center}
\end{table}

\section{Methodology}\label{sec:model}

\subsection{Picking out satellite candidates}\label{sec:sat_cand}

We use the level-2 merger trees of the six Aquarius halos to determine
which dark matter halos can host luminous satellite galaxies based on
several criteria.  These again reflect our assumptions about when a
subhalo is able to accrete gas from the IGM, cool its gas, and form
stars, following methods developed in e.g. \citet{Madau2008b, Koposov2009, Busha2010}.  We only consider subhalos that end up within the virial radius of their host halos (see Table~\ref{tab:aq_halos}).
 
First, we require that subhalos must grow larger than a minimum
$v_{max}$ of 15 km/s before star formation can take place, 
and also that this happens before the subhalo is reionized.
This corresponds roughly to the virial
temperature where atomic hydrogen is no longer able to cool ($\sim
10^4$ K).  We do not explicitly include H$_2$-cooling ``minihalos'' as
sites for low-mass star formation before reionization, but recognize
that such halos may have been the sites of the very first star
formation, whose metals seeded the next generation of lower-mass,
long-lived stars \citep{Bromm1999, Bromm2003b,Yoshida2004,
Bromm2009}.

Once a subhalo enters a reionized region (as described in
Section~\ref{sec:reion_cal}), the minimum $v_{max}$ needed to sustain
star formation increases, reflecting both the increased temperature of
the IGM to a few times $10^4$ K due to photoionization, and the
possibility of photoevaporation of gas out of small halos
\citep{Barkana1999, Iliev2005}.  \citet{Gnedin2000} found the
filtering mass to correspond to the scale at which the gas fraction in
halos is significantly reduced due to reionization.  Gnedin's
expression, however, does not take into account that only the cold gas
remains available for star formation; \citet{Munoz2009} and
\citet{Busha2010} argue that only halos that grow to a viral
temperature $\sim 10^5$ K will be able to sustain star formation after
reionization.  Accordingly, we only let star formation continue
post-reionization in subhalos whose $v_{max}$ is greater than $50
\mbox{ km s}^{-1}$.  This represents a rather abrupt reionization
effect; we also run a model where this suppression is weaker, with the
threshold set at $30 \mbox{ km s}^{-1}$.

In addition, we assume that once a satellite starts interacting with
the main halo, its gas is stripped and no further star formation takes
place.  Thus, there are two important redshifts associated with each
satellite: $z_{re}$ and $z_{{\rm infall}}$, the times when it is
reionized and falls into the main halo, respectively.

In order to account for tidal stripping of satellites, which has been
proposed as another important mechanism to resolve the missing
satellite problem \citep{Kravtsov2004b}, we tag the 1\% most bound
particles of the subhalos as star-carrying when they are accreted onto
the main halo. We then track all these tagged particles to redshift zero, and determine the final luminosity of a satellite based on its remaining tagged particles. In this way, we are able to take into account tidal stripping of satellites and other dynamical effects, under the assumption that the stars are found in the bottom of 
the potential well.

\begin{figure*}
 \begin{center}
  \includegraphics{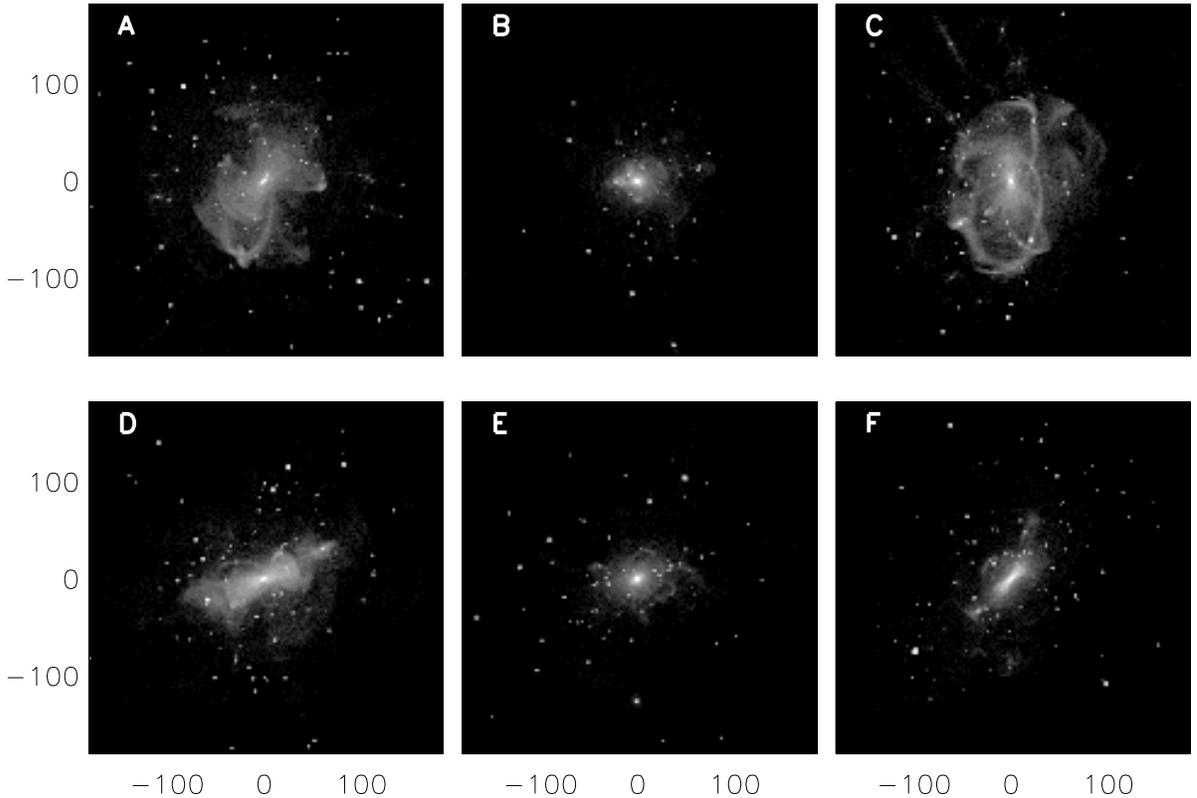}
  \caption{Projected distribution of all dark matter particles that are tagged with stellar populations from infalling subhalos, at $z=0$. The scale is in kiloparsecs. Both the concentrated, bright satellites and the shredded satellites that make up the diffuse accreted stellar halo and streams are visible.}
  \label{fig:halopic}
 \end{center}
\end{figure*}

Of course, a number of infalling systems will be completely disrupted, and contribute to the accreted stellar halo. In Fig.~\ref{fig:halopic}, we show the projected distribution of all the dark matter particles that were tagged as stars on infall, in reionization scenario 1. Surviving satellites here appear as bright, concentrated dots, while the particles from the shredded systems make up the diffuse halo. Since the accreted stellar halos and streams forming in the
Aquarius simulations have been studied extensively in \citet{Cooper2010} and
\citet{Helmi2011}, we will not discuss them further here. We note,
however, that although our model is very simple, it still reproduces
most of the features seen in Fig. 6 of \citet{Cooper2010} -- we take this as justification that our simple model still reasonably captures the relevant processes.

We determine satellite candidates for all four of our reionization
calculations.  In addition, we do the same for four ``instantaneous''
redshift models, where we set the reionization redshift equal to the
mean reionization redshift in the entire box. \citet{Alvarez2009}
showed that the distribution of reionization redshifts for Milky
Way-sized halos is nearly similar to the global spatial distribution,
making the global mean an appropriate choice.  Comparing the resulting
satellite populations from the patchy and instantaneous reionization
models then allows us to isolate any effects of reionization not being
spatially uniform.

\subsection{Assigning luminosities to halos}

In order to compare our satellite candidates to the observed Milky Way
population, we need to assign luminosities to the subhalos. There are
many different procedures for doing this; see e.g., \citet{Koposov2009}
for a discussion.  Here, we adopt a simple model where the star
formation rate is proportional to the mass, that is $M_{star} = \int
\epsilon \times f_{gas} \times M_{DM} \, dt$, where $\epsilon =
1/\tau$ is a constant that sets the star formation efficiency
(i.e., $\tau$ can be considered a timescale for which a subhalo would
convert all of its gas into stars). We take $f_{gas} \times \epsilon =
0.08 \times 10^{-10} \mbox{yr}^{-1}$, tuned both so that we get a
reasonable satellite population, but also so that the mass of the
accreted stellar halo (i.e., the total mass of satellites that fall in
and are completely shredded) falls in a mass range around $\sim 10^9
M_{\odot}$ \citep{Bell2008, Tumlinson2010}. This second criterion is 
insensitive to the
reionization prescription, because the mass of the stellar halos is dominated by a few significant progenitors, while the smaller systems 
affected by reionization contribute a small fraction of the total mass. While in principle we could increase this product to get a better fit for the more extreme reionization scenarios, this would then lead to very massive stellar halos.

Our model is deliberately simple because we wish to assess and compare
the impact of different reionization assumptions without getting
addressing uncertainties in the baryonic physics.  While we do not
explicitly model effects like the impact of supernova feedback on the
satellites, we assume that
the typical effect is captured by setting
$\epsilon$ (i.e., the star formation efficiency) appropriately. (For an example of a more extensive approach in modeling the baryonic content of dark halos in cosmological simulations, we refer the reader to \citealt{Font2011}.)

Having determined the stellar mass formed, we then assign each stellar
population an associated present-day luminosity, using the stellar
population synthesis (SPS) models of \citet{Bruzual2003}, assuming a
Salpeter initial mass function, and a metallicity of $Z=0.0001$ (the
lowest considered in their model, and which captures the typical range
of the ultra-faint satellites we are most interested in, e.g.,
\citealt{Kirby2008b,Simon2010a,Norris2010a}).  We then track the
particles tagged with stellar populations to redshift zero, and so
determine the present-day luminosities of the satellite candidates
picked out as described earlier.

As an alternative to the simple parametric method described above, we
also try the abundance matching technique developed by
\citet{Busha2010} to assign luminosities to dark matter halos.  This
is an extrapolation of the fit of \citet{Blanton2005} based on
galaxies in the SDSS (down to $M_r = -12.375$) to fainter magnitudes,
by a power law:

\begin{equation}
M_V - 5 \log h = 18.2 - 2.5 \log \left[ \left( \frac{v_{max}}{1 \mbox{km s}^{-1}} \right) ^{7.1} \right] .
\label{eq:ab_match}
\end{equation}

\noindent The appeal of the abundance matching procedure is that there
are no parameters to tune, using a model derived from statistics of
observed galaxies.  While it is an extrapolation at UFD luminosities,
\citet{Busha2010} showed that when applied to the Via Lactea subhalos,
the abundance matching model did well in reproducing the observed
luminosity function (but see \citealt{Boylan-Kolchin2011} for a discussion of potential problems with abundance matching for the MW dwarfs).  For purposes of comparison, then, we also separately 
assign magnitudes to our subhalos using eqn.~\ref{eq:ab_match}.  In
order to select an appropriate $v_{max}$ for the calculation, we use
the peak $v_{max}$ of the subhalos during the time they were able to
form stars, as defined in Section~\ref{sec:sat_cand}.  For most
subhalos, this either corresponds to the time of reionization or time
of accretion.

\begin{figure}
 \begin{center}
  \includegraphics[width=8.5cm]{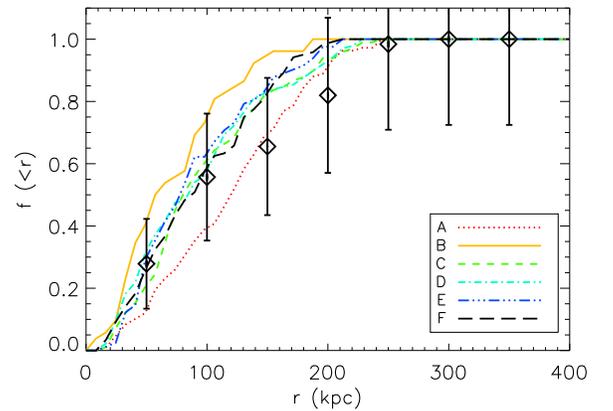}
\caption{Radial distribution of satellites, as a fraction
  of total number. The points with error bars show the observed Milky
  Way distribution, where non-SDSS satellites are weighted by a factor
  of $f_{DR5}=0.194$ of the total to account for incomplete sky
  coverage. The theoretical estimates are for reionization Model 1 with
  strong suppression, but the trends are similar for the other
  models. In particular, all halos have a radial satellite
  distribution consistent with that observed around the Milky Way,
  with the possible exception of the B halo, whose satellites are
  somewhat closer to the main halo.}
\label{fig:rdist}
 \end{center}
\end{figure}

As some basic tests of our model, we check how well it reproduces various features of the observed Milky Way dwarf population, such as their radial distribution, and observed mass-to-light ratios. Fig.~\ref{fig:rdist} shows the radial distribution of satellites in our simulation, as a fraction of total number. We find all six halos to have a satellite distribution consistent with what is observed around the Milky Way, with the possible exception of the B halo, which has a larger fraction of satellites somewhat closer in.

\begin{figure*}
 \begin{center}
  \includegraphics{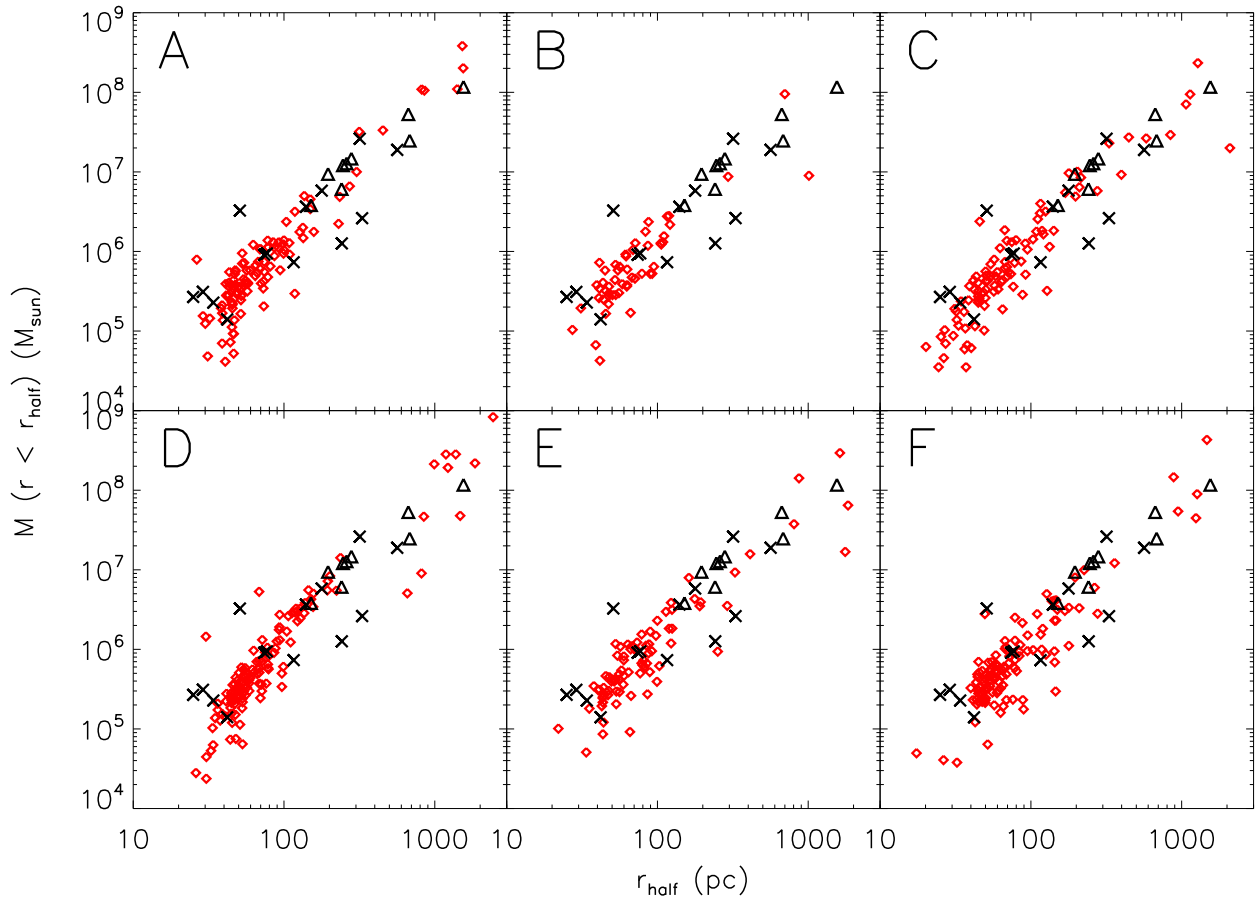}
  \caption{Mass within the estimated half-light radius versus
  half-light radius for the simulated satellites (red diamonds) for
  the six halos, in reionization scenario 1, with the higher suppression ($v_{max} = 50 \mbox{ km s}^{-1}$). For the simulation points, the half-light radius is estimated from the luminosity via an empirical relation, and the mass is measured by fitting a NFW profile. Black triangles show the
  observed values in the Milky Way classical dSphs, while the crosses show the values for the UFDs.}
  \label{fig:mrhalf}
 \end{center}
\end{figure*}

As for mass-to-light ratios, we note that subhalo masses in a simulation cannot be directly
compared to observed dwarf galaxy masses. However, \citet{Walker2009,
Walker2010} and \citet{Wolf2010} show that the mass within the half-light radius is well
constrained for the Milky Way dwarfs. We thus estimate the half-light
radius of our satellites from their assigned luminosities by fitting a
power law to the data tabulated in \citet{Walker2010}, and measure the
mass within the half-light radius by fitting an NFW profile
\citep{Navarro1997}. The result is shown in Fig.~\ref{fig:mrhalf}: the
red points are simulated satellites, and the black crosses (UFDs) and triangles (classical dSphs)
show the data from the Milky Way dwarf galaxies. Our satellite
properties are in good agreement with the observed values, and
recover the observed relation between $M$ and $r_{0.5}$. The points
shown are for satellites with the Model 1 reionization history;
driving reionization to earlier times yield the same relation, but
fewer satellites overall, and fewer in the intermediate $r_{0.5}$ (or
luminosity) range. We conclude that our simple model is overall
sufficient for reproducing the observed trends.

\section{Results}\label{sec:res}

\subsection{Spread in reionization redshifts}
\label{sec:reion_res}

A major novel factor in our work is that the reionization redshift of a halo is completely determined by its environment. This has two main 
consequences: first, that the six main halos will reionize at different times according to their 
locations in the Millennium-II box.  But additionally, the subhalos of each of these main halos are not required to reionize at the same time as their main halo - their reionization times are also set by their particular path through the main box. Here, we explore the distribution in reionization redshifts for the subhalos, and how this 
influences the resulting number of potential satellites.

Table~\ref{tab:reion_halo} summarizes the first effect -- that the six Aquarius host halos do not reionize at the same times in the box. In particular, the F halo consistently reionizes later than the box mean, while the C halo reionizes significantly earlier. The other four halos are closer to the mean of the box (and thus slightly earlier than the median; also see Table~\ref{tab:reion_mod}).

\begin{table}
 \begin{center}
 \caption{Reionization times of the main halos in the four models}
 \label{tab:reion_halo}
  \begin{tabular}{c c c c c c c c}
   \hline \hline
  Model No. &  A     &  B     &  C    &  D    &  E    &  F   & Box Mean  \\ \hline 
  1         &  9.2   &  9.9   & 12.0  & 9.3   & 8.5   & 8.1  &   9.31    \\	          
  2         &  11.5  &  11.2  & 13.7  &	10.7  & 10.0  & 9.5  &  10.71    \\	          
  3         &  12.6  &  12.4  & 16.3  &	12.0  &	11.8  &	10.9 &  11.97    \\	          
  4         &  14.0  &  14.3  & 17.1  &	13.5  &	13.6  &	12.5 &  13.50    \\ \hline     

  \end{tabular}
 \end{center}
\end{table}

\begin{figure}
 \begin{center}
  \includegraphics[width=8.5cm]{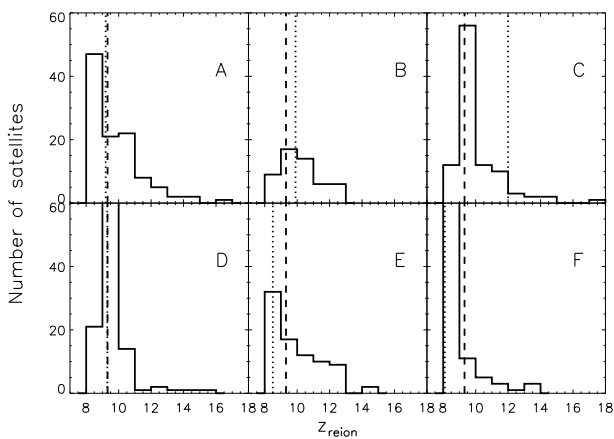}
 \end{center}
 \caption{Distributions of reionization redshifts of the satellite
 candidates of the six halos, here shown for Model 1 and the higher ($v_{max} = 50 \mbox{ km s}^{-1}$) suppression threshold. The dashed lines
 correspond to the mean reionization redshift across the entire box in
 Model 1, while the dotted lines show the reionization redshifts of
 the six main halos. Note the tails to higher redshifts.}
 \label{fig:sat_dist}
\end{figure}

As for the second effect: Fig.~\ref{fig:sat_dist} shows the distribution of satellite reionization redshifts for the six halos, here in the case of scenario 1. (The distributions for the other three models show similar characteristics, but with fewer satellites and narrowing 
distributions as the models 
become more efficient.) Several interesting trends are evident: First, there {\it is} a distribution -- while some of them are strongly peaked (C, D and F in particular), all show a range of redshifts. Second, the peak of the distribution does not necessarily correspond to the reionization redshift of their host halo (the dotted lines). 

The C halo is the most extreme in this regard -- the peak of the satellite distribution is near the mean of the box, while the host reionized significantly earlier. Essentially, this happens because the 
location 
of the C halo is near the edge of one of these early-reionized regions in the large box (cyan in Fig.~\ref{fig:reion_map}), responsible for the early reionization of the C halo itself. Many of its satellites, however, fall in from a different direction than the bubble front, and end up with a reionization distribution peaking around the box mean. Note that Fig.~\ref{fig:sat_dist} shows only systems that survive as self-bound satellites. If we were to also include systems that are shredded and 
are part of the diffuse main halo by redshift zero, we would see a larger number of subhalos reionizing around the same time as the C main halo.

\begin{figure}
 \begin{center}
   \includegraphics[width=8.5cm]{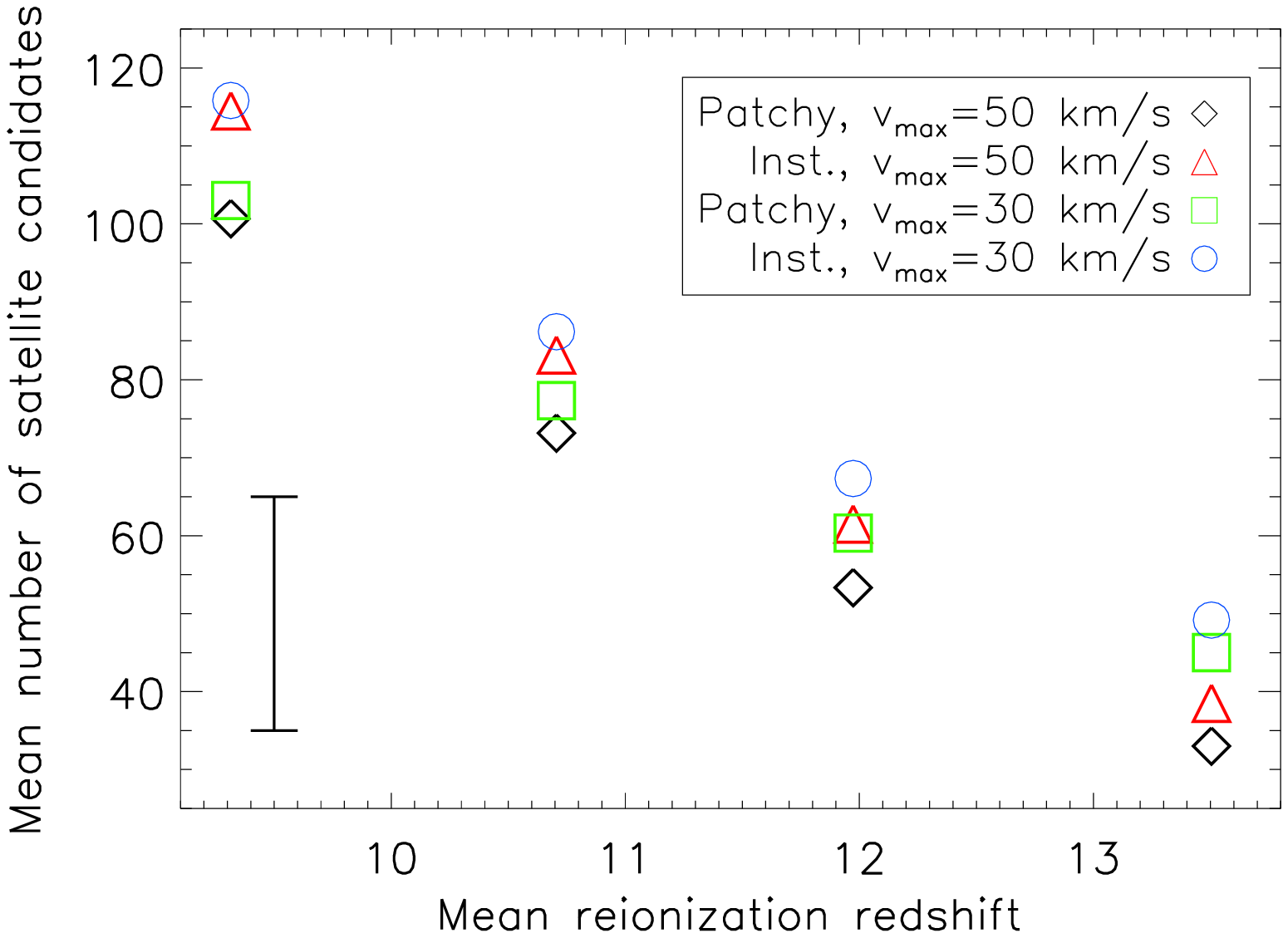}
 \caption{Mean number of satellite candidates for the six halos, for
 our four different reionization models, with varying assumptions:
 Either strong suppression at reionization, affecting all halos with
 $v_{max} < 50$ km s$^{-1}$, or a weaker suppression at 30 km
 s$^{-1}$.  In addition to the patchy cases, we also show
 calculations where reionization instead happened
 instantaneously at the mean redshift of each patchy model.  While the
 trends with reionization model and suppression $v_{max}$ are similar
 for the six halos, there is substantial halo-halo scatter in the
 actual number of satellite candidates.  This is indicated by the error
 bar (typical rms $\sim$ 15 satellites). }
 \label{fig:sat_cand}
 \end{center}
\end{figure}

Having established that there is indeed a distribution of satellite reionization redshifts, the next question is whether this leads to a different satellite population between patchy and instantaneous models. Fig.~\ref{fig:sat_cand} 
considers the number of satellites produced in each model for different sets of assumptions: either patchy or instantaneous, and a suppression $v_{max}$ of either 30 or 50 km/s. We see that regardless of cutoffs and patchy vs. instantaneous, as reionization is pushed to earlier times, there is a strong decline in the number of satellites. This overall trend is similar to what was found in previous studies \citep{Munoz2009, Tumlinson2010, Busha2010}. For a given reionization model, we do see
additional differences, however.  For either choice of suppression $v_{max}$, the
instantaneous reionization models systematically make 10-20\% more
satellites than the corresponding patchy models; in the more extended
models this effect dominates the difference between stronger and
weaker suppression.  As the reionization sources become more efficient
and reionization occurs earlier, the patchiness is less important,
and the variation in suppression $v_{max}$ becomes the dominant effect in this scenario.

Note that the numbers shown in Fig.~\ref{fig:sat_cand} are averages; as shown in Fig.~\ref{fig:sat_dist} there is a substantial spread between the actual number of satellites between the different halos.  The error bar indicates the typical
rms, $\pm \sim 15$ satellites.  However, halos A-E all show more
satellite candidates in the instantaneous scenarios, while the F halo
has about 10\% fewer.  For five out of the six halos, then, the main
effect of patchy reionization is that the number of satellite
candidates is reduced by 10-20\% compared to the instantaneous models. This can be understood in the context of Fig.~\ref{fig:sat_dist}, where the F halo is the only halo whose majority of satellites reionize later than the mean.

In practice, the difference between the patchy and instantaneous case in Fig.~\ref{fig:sat_cand} is due to a combination of two effects: that the global mean is not necessarily the best instantaneous redshift to choose for a given halo, as well as not capturing the spread in reionization times. However, we do not find that using e.g. the host halo reionization redshift, or the median satellite reionization redshift, leads to better agreement in each case. While we could recover the number of satellites in each case by tuning the instantaneous reionization redshift, the 
value of such an exercise is limited: the only way to determine that appropriate redshift would be to already know the result from the patchy model. As such, this illustrates the main problem with using instantaneous and/or spatially homogeneous reionization models. Depending on the application, the fact that there is a distribution of satellite reionization redshifts may be important to the results -- and this distribution cannot be captured in a homogeneous model, nor can the appropriate instantaneous approximation be recovered without knowing the patchy result.


\subsection{Luminosity Functions}

Having explored the reionization distributions in our simulations, we here discuss the resulting luminosity function of our simulated satellites. 
In order to compare them to the population of
observed dwarfs, however, we first need to correct the observational sample of
dwarf galaxies (data taken from \citealt{Martin2008, Wadepuhl2011}) for incompleteness and biases to establish a luminosity
function.  We follow the method outlined in \citet{Koposov2008} and
\citet{Tollerud2008}, which employed models for the completeness of
the SDSS Data Release 5 (DR5) to quantify the detection efficiency and
thus adjust the observed luminosity function.  In addition, we only
include the simulated satellites that would be detectable according to
the DR5 criteria:
\begin{equation}
 r_{max} = \left(\frac{3}{4\pi f_{DR5}} \right)^{1/3} 10^{\left(-0.6 M_V -5.23\right)/3} \mbox{ Mpc},
 \label{eq:rmax}
\end{equation}
\noindent where $f_{DR5} = 0.194$ is the fraction of the sky covered
by DR5.  That is, simulated satellites with $ r > r_{max}(M_V)$ are not
included in the plots in this section. In each case, $r_{max}$ is calculated with respect to a point 8 kpc from the center of the main halo.

\begin{figure*}
 \begin{center}
  \includegraphics{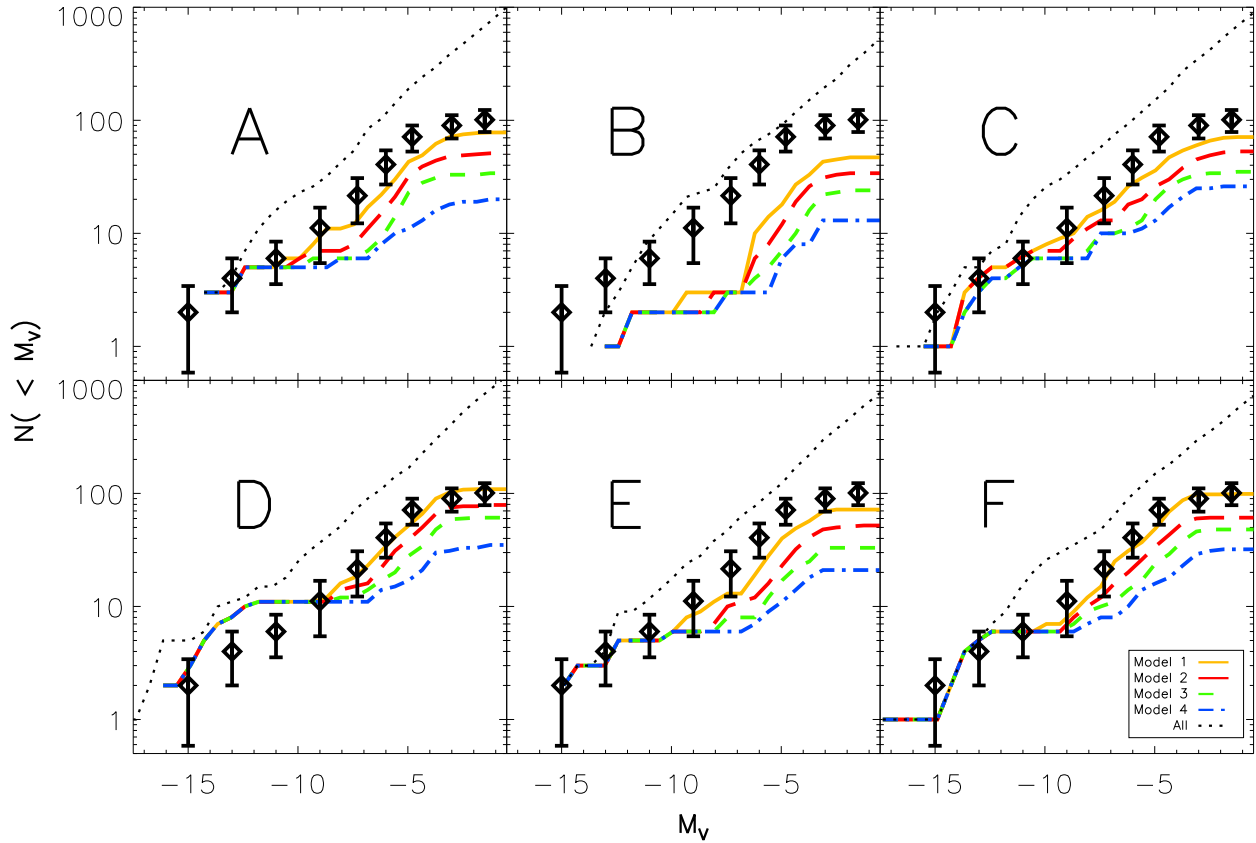}
  \caption{Cumulative luminosity functions of the simulated satellite populations of the six halos, using the abundance matching technique to assign luminosities. The colored curves show the results for the four patchy reionization models, with strong suppression at reionization ($v_{max}=50 \mbox{ km s}^{-1}$). The black points with error bars show the observed Milky Way satellite luminosity function, applying the completeness correction of \citet{Koposov2008} and \citet{Tollerud2008}. The dotted line shows what the luminosity function for all the subhalos would be if we ignored all baryonic effects, and used the same abundance matching extrapolation to assign luminosities to all subhalos in each case.}
  \label{fig:lumfun}
 \end{center}
\end{figure*}

Figure~\ref{fig:lumfun} shows the resulting cumulative luminosity
function for the simulated satellite population of all six halos, for
the four patchy reionization models and with $v_{max}=50 \mbox{ km
s}^{-1}$.  The luminosities shown are those obtained with the
abundance matching technique (eqn.~\ref{eq:ab_match}), but the
resulting luminosity functions are very similar to those using our
simple SPS-based model.  (However, there is substantial scatter between the methods on a
subhalo-subhalo basis, i.e., rms $\sim 1$ mag.) The dotted line shows the resulting luminosity function when applying the abundance matching prescription to all subhalos in each simulation that 
lie within the virial radius by redshift $z=0$, without assuming any suppression due to baryonic processes. As such, it gives a good illustration of the 
missing satellite problem: without some process such as reionization to suppress star formation and reduce the number of luminous satellites, the six Aquarius halos would predict between 500-1000 satellites around Milky Way-like galaxies. 

We see that while satellites brighter than $M_V \sim -10$ are insensitive to the
reionization model, the total number and the number of faint satellites change dramatically with more efficient reionization. This is consistent with the results of \citet{Font2011}, who found that photoheating is mainly important for suppressing star formation in satellites analogous to ultra-faint dwarfs. In our simulations, only Model 1, and in some cases Model 2, can match the faint end of the luminosity function. In fact, there is a factor of 3-4 difference between the total number of satellites predicted between the different reionization models for a given halo. This behavior may ultimately allow for constraining the Milky Way reionization epoch with the faintest satellites.

Another striking feature of Fig.~\ref{fig:lumfun} is the scatter
between the six halos -- there is a factor of $\sim3$ difference
between the B and D halos, which have the fewest and most satellites,
respectively. Thus, the halo-halo scatter is of the same order as the
differences in the impact of the various reionization models 
on the faint end.  In order
to break this degeneracy, it will be necessary to understand the
general distribution of substructure around Milky Way-like halos,
requiring a statistical sample of halos at sufficient
resolution. Since the bright end of the luminosity function is not
affected by the reionization history, however, it may be possible to
break the degeneracy by constraining the halo-halo scatter using the
brighter satellites, and use the fainter satellites to constrain the
reionization epoch of the Milky Way.

We find that only the F halo hosts a satellite as bright as the
Magellanic Clouds. This is consistent, however, with recent studies
based on SDSS DR7 on the abundance of close, bright satellites around
Milky Way-like galaxies \citep{Liu2011, Boylan-Kolchin2010b,Busha2010b, Guo2011,
Tollerud2011}, and the possibility that the Clouds have entered the
Milky Way halo only recently \citep{Besla2007,Besla2010}.
Hence, we do not consider the mismatch at the brightest
magnitudes to be problematic, especially given the simplicity of our
star formation model.

Similar to the brighter satellites, the total stellar masses in the accreted halo of shredded,
infalling satellites do not change significantly with reionization
model.  This is consistent with the results of \citet{Cooper2010}, who
studied in detail the building of accreted stellar halos in the
Aquarius simulations.  They found that the significant progenitors are
similar in mass to the brightest Milky Way dwarf spheroidals, with
each halo assembled from less than five such objects.

\begin{figure*}
 \begin{center}
  \includegraphics{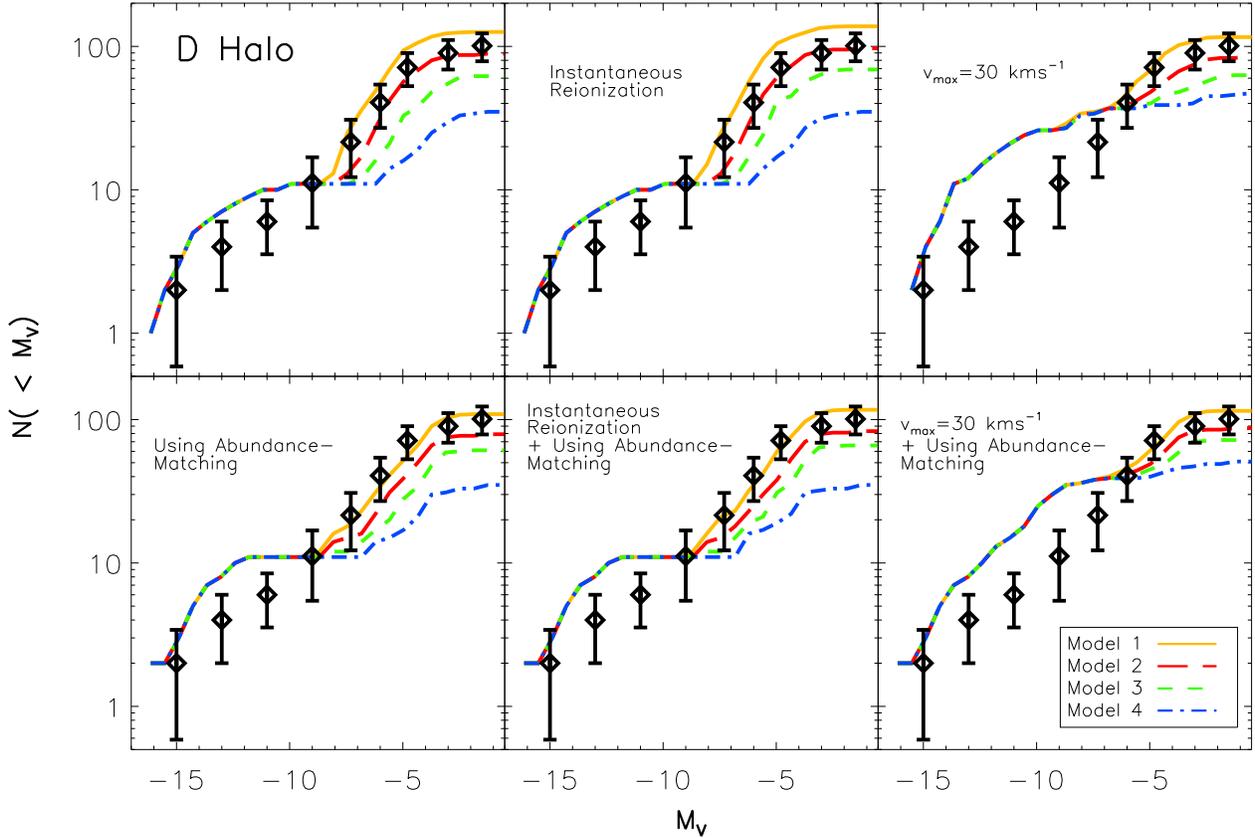}
  \caption{Comparisons of the different models, here illustrated by the resulting luminosity function for satellites of the D halo. Upper left: same as Fig.~\ref{fig:lumfun}: the patchy reionization model, with luminosities assigned by the SPS method, for $v_{max}=50 \mbox{ km s}^{-1}$. Upper middle: instantaneous reionization at the mean reionization redshift of the model. Upper right: patchy, with threshold for suppressing star formation lowered to $v_{max}=30 \mbox{ km s}^{-1}$. Lower panels: same as upper, but with luminosities assigned by the abundance matching technique, rather than the SPS method. }
  \label{fig:lumfun_comp}
 \end{center}
\end{figure*}

To highlight some of the differences between the various model
assumptions, Fig.~\ref{fig:lumfun_comp} shows the luminosity function
of the D halo satellites for some parameter choices.  The upper panels
show the results using patchy reionization and $v_{max}=50 \mbox{ km
s}^{-1}$, instantaneous reionization and $v_{max}=50 \mbox{ km
s}^{-1}$, and patchy reionization and $v_{max}=30 \mbox{ km s}^{-1}$
respectively.  The three lower panels show the same, but with
luminosities determined by abundance matching
(eqn.~\ref{eq:ab_match}).

While the models with post-reionization star formation suppressed at
$v_{max}=50 \mbox{ km s}^{-1}$ give a good match to the observed
luminosity function, when we decrease the suppression threshold to $30
\mbox{ km s}^{-1}$ none of the models fit the bias-corrected
luminosity function well, despite the overall number of satellites only being slightly increased. This holds true both for the abundance matching-derived
luminosities, and the best-fit SFR/SPS model.  In particular, even if
the star formation efficiency is decreased to give the correct number
of satellites overall, the shape still does not match the observed
luminosity function.  It is possible, however, that this lower
suppression model could still be a good match, if we explicitly
consider other processes like supernova feedback \citep{Li2010}.

\begin{figure}
 \begin{center}
  \includegraphics[width=8.5cm]{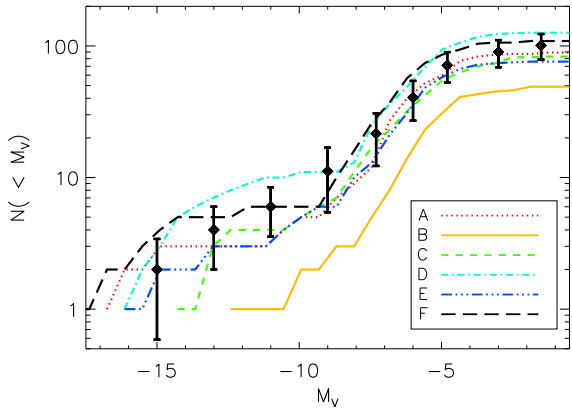}
  \caption{Illustrating halo-halo scatter: Overplotting the luminosity functions of the satellite populations of all the six halos, here shown for reionization model 1 (yellow line in Figs.~\ref{fig:lumfun} and \ref{fig:lumfun_comp}).}
  \label{fig:all_halos}
 \end{center}
\end{figure}

Compared to patchy reionization, the instantaneous models generally yield more satellites.
While only the D halo is shown here as an illustration, as discussed in Section~\ref{sec:reion_res} this holds true for all of the Aquarius halos, except for the F halo. It is interesting to note, however, that the differences between different reionization models, luminosity prescriptions, etc., are in general dominated by the halo-halo scatter. That is, the differences in the satellite populations of the six halos for any given model are as large as the differences between the range of models we are exploring. As an example, Figure~\ref{fig:all_halos} shows the satellite luminosity functions for all six halos overplotted, here for reionization model 1 and the SFR luminosity prescription. While the B halo, which is the clear outlier, is also the lowest mass halo in this simulation suite, the uncertainty of where the Milky Way itself fits in this distribution makes direct comparisons to constrain models difficult.

\section{Discussion and Conclusions}\label{sec:conc}

In our models, we find that the population of faint satellites is
highly sensitive to the reionization model.  This agrees with the
results of \citet{Busha2010, Font2011}, but is at odds with several earlier
studies in the literature \citep{Somerville2002, Kravtsov2004b}.  The
difference is largely due to two factors. First, the previous work
that focused on the range of the classical dwarf galaxies of the Milky
Way found little reionization dependence; our results agree to the
extent that the bright end of our luminosity function (classical dSph
regime with $10^{5}<L<10^{8}\,L_{\odot}$) does not change either with large scale
(different reionization model) or small scale (patchy versus
instantaneous) changes in reionization.  The difference at the faint
end is explained by the various assumptions about how reionization
influences the available supply of cold gas in small halos -- our
model assumes an abrupt end to star formation in halos smaller than a
cutoff $v_{max}$, whereas the filtering mass formalism will lead to a
gradual suppression in star formation, and a much weaker reionization
dependence.  In that sense, we have chosen a maximum effect, and the
results of our study can be viewed as an upper limit to the impact of
patchiness of reionization.

We also note that since all but the largest subhalos in our simulation
finish their star formation around reionization, our results are
consistent with some ultra-faint dwarf galaxies having formed around
the epoch of reionization in atomic cooling halos (but see e.g.,
\citealt{Bovill2009}).  Given the cutoff we set at the atomic cooling
threshold, our model does not predict objects much fainter than the
faintest dwarf galaxies already observed, although this could
potentially change if H$_2$ cooling at high redshift were included.

Our main conclusions can then be summarized as follows:
\begin{itemize}

\item Patchy reionization models yield a distribution in satellite reionization redshifts, which leads to a 10-20\% difference in the number of satellites they predict compared to instantaneous models. This effect can result in more or fewer satellites depending on the overall environment and merger history of the halo, and as such there is no general way to correct for it in a homogeneous model. 

\item The overall number of satellites depends sensitively on the reionization model -- we find a factor of 3-4 difference between the earliest and latest reionization model. The difference is entirely in the number of fainter satellites, while the brighter end of the luminosity function corresponding to the classical dSph regime is unaffected.

\item There are large halo-halo variations in the number of luminous
satellites for a Milky Way sized halo -- we find a factor of 2-3
difference between the halo with fewest and most satellites, comparable to the spread between the different reionization models.  Hence,
in order to use the Milky Way satellites to constrain reionization, a
much larger sample of halo simulations is required.  Given the spread
between the six halos in the Aquarius sample, the direct comparison of
one or a few halos to the observed Milky Way will only yield
conclusive results once we have a more detailed understanding of how our
Galaxy fits into this distribution.

\end{itemize}

Progress can be made in several directions.  On the observational
side, future and ongoing surveys like SkyMapper, PanSTARRS and LSST
\citep{Keller2007, Kaiser2002, Ivezic2008} will cover the entire sky
at the depth of SDSS or deeper, reducing the uncertainty in the number
of observed dwarf galaxies due to the incomplete sky coverage of
SDSS.  Follow-up studies of new and existing dwarf galaxies can also
provide more observational constraints, which can help discriminate
between the different theoretical models that at present all fit the
observed luminosity function.

On the simulation side, the present work has shown that in a six-halo
sample, there is a large degeneracy between the significant
halo-to-halo scatter and the results of effects like
reionization. This level of cosmic variance will need to be quantified
more accurately with a larger set of simulated halos. Only then can
the origin and evolution of the faintest galaxies in the universe, as
well as their role in building larger galaxies be more thoroughly
understood.

\section{Acknowledgments}
We thank the Virgo Consortium for access to the Aquarius simulations for
this project, and Matt Walker and Daniel Eisenstein for helpful
discussions. We also thank the anonymous referee for helpful comments that improved the structure of the manuscript. A.F. acknowledges support of a Clay Fellowship
administered by the Smithsonian Astrophysical Observatory.

\end{document}